\documentclass[aps, prl, twocolumn, groupedaddress, showpacs, showkeys]{revtex4}
\usepackage{graphicx}
\usepackage{epstopdf}
\usepackage{dcolumn}
\usepackage{amsmath,amssymb}

\def\bx{{\mathbf{x}}}
\def\by{{\mathbf{y}}}

\newcommand{\beq}{\begin{equation}}
\newcommand{\eeq}{\end{equation}}
\newcommand{\bea}{\begin{eqnarray}}
\newcommand{\eea}{\end{eqnarray}}
\newcommand{\ba}{\begin{array}}
\newcommand{\ea}{\end{array}}
\newcommand{\bit}{\begin{itemize}}
\newcommand{\eit}{\end{itemize}}
\newcommand{\nn}{\nonumber}
\newcommand{\eq}{Eq.~}

\newcommand{\fig}{Fig.~}

\def\math{\mathsurround 0pt}
\def\oversim#1#2{\lower.5pt\vbox{\baselineskip0pt \lineskip-.5pt
        \ialign{$\math#1\hfil##\hfil$\crcr#2\crcr{\scriptstyle\sim}\crcr}}}

\def\lsi{\raise0.3ex\hbox{$<$\kern-0.75em\raise-1.1ex\hbox{$\sim$}}}
\def\gsi{\raise0.3ex\hbox{$>$\kern-0.75em\raise-1.1ex\hbox{$\sim$}}}
\newcommand{\lsim}{\mathop{\lsi}}
\newcommand{\gsim}{\mathop{\gsi}}

\newcommand{\de}{\mathrm{d}}

\begin{document}

\title{\vspace{-12pt}
Onset Transition to Cold Nuclear Matter from Lattice QCD with Heavy Quarks} 
\author{M.~Fromm, J.~Langelage, S.~Lottini, M.~Neuman, O.~Philipsen} 
\affiliation{Institut f\"ur Theoretische Physik, Johann-Wolfgang-Goethe-Universit\"at, \\
                   Max-von-Laue-Strasse~1, 60438 Frankfurt am Main, Germany}

\date{\today}

\begin{abstract}
Lattice QCD at finite density suffers from a severe sign problem, which has
so far prohibited simulations of the cold and dense regime. Here we study the onset of nuclear matter employing a three-dimensional
effective theory derived by combined strong coupling and hopping expansions, which is valid
for heavy but dynamical quarks and has a mild sign problem only.
Its numerical evaluations agree between a standard Metropolis and complex Langevin algorithm, where the latter is free of the sign problem. Our continuum extrapolated data approach a first
order phase transition at $\mu_B\approx m_B$ as the temperature 
approaches zero. An excellent description of the data is achieved by an analytic solution in the strong coupling limit.
   
\end{abstract}


\maketitle


QCD at zero temperature is expected to exhibit the so-called 
silver blaze property: when a chemical potential for baryon number $\mu_B$ 
is switched on in the grand canonical partition function, 
initially all observables should be 
completely independent of $\mu_B$. 
This changes abruptly once the chemical potential exceeds a critical 
value $\mu_{Bc}$, for which the baryon number jumps from zero to a finite value
and a transition to a condensed state of nuclear matter takes place. 
The reason for this behavior is 
the mass gap in the fermionic spectrum, where the nucleon mass $m_B$ 
represents the lowest baryonic energy that can be populated 
once $\mu_{B}\approx m_B$. While this picture is easy to see in terms 
of energy levels of nucleons in a Hamiltonian language, 
it is elusive in the fundamental formulation of QCD thermodynamics in 
terms of a path integral. There, chemical potential enters through the 
Dirac operators of the quark fields, and hence all Dirac  
eigenvalues are shifted for any value of $\mu_B$. 
The silver blaze property thus requires some exact cancellations 
for $\mu_B<m_B$. 

An analytic derivation of the silver blaze property from the path integral 
exists only for the related case of finite isospin chemical potential $\mu_I=\mu_u=-\mu_d$ \cite{cohen}, where Bose-Einstein condensation of pions sets in
at $\mu_I=m_\pi/2$. A numerical demonstration of the effect by means of lattice QCD
has so far been impossible due to the so-called sign problem. 
For finite baryon chemical potential the action 
becomes complex, prohibiting its use in a Boltzmann factor for Monte 
Carlo approaches with importance 
sampling. Several approximate methods have been devised to circumvent this problem.
These are valid in the regime $\mu\lsim T$, where they give consistent results (for a recent review see \cite{review}). However, the cold and dense region of QCD has so far been inaccessible to 
lattice simulations. A method avoiding importance sampling
is stochastic quantization, where expectation values are obtained from equilibrium distributions of
stochastic processes governed by a Langevin equation \cite{langevin}. 
While this works for several models with a sign problem \cite{aarts2},
it is not generally valid for complex actions \cite{dh}.
Using Langevin dynamics, the silver blaze property has been numerically demonstrated 
for the Bose condensation of complex scalar fields \cite{aarts_bose}.
This was recently reproduced using a worm 
algorithm on the flux representation of the complex action,
which is free of the sign problem \cite{gatt_bose}. 

In this work we show that cold and dense lattice QCD is accessible within a 3d effective theory of Polyakov loops, which has been derived from the full lattice theory with Wilson fermions by means of strong coupling
and hopping parameter expansions \cite{efft_ym,efft_kappa}. 
The pure gauge part reproduces
the critical temperature $T_c$ of the deconfinement transition
in the continuum limit to a few percent accuracy \cite{efft_ym}. 
The theory was extended to include heavy but dynamical Wilson quarks.
The sign problem of the resulting effective theory being under full control,
the finite-temperature deconfinement transition, including its surface of
endpoints, was located for {\it all} chemical potentials.
The critical quark mass corresponding to $\mu_B=0$ was again found to quantitatively agree with full 4d Wilson simulations \cite{efft_kappa}. 
The current restriction to large quark
masses ensures the validity of the hopping expansion, we comment on possible extensions later.  

The lattice QCD partition function with Wilson gauge action $S_g[U]$ and $f=1,\ldots,N_f$ quark 
flavours with Wilson fermion matrix $Q(\kappa_f,\mu_f)$ 
can be written as
\bea
	Z&=&\int [\de U_\mu]\prod_f\det[Q]e^{-S_g[U]}
		=\int [\de W]\,e^{-S_\text{eff}[W]}\;;\nn\\
	S_{\mathrm{eff}}&=&S^s_\text{eff}+S^a_\text{eff}\;;\quad
		S^s_{\mathrm{eff}}[W]=-\sum_{i=1}^\infty\lambda_iS_i^s[W]\;;\\
	S^a_\text{eff}[W]&=&2\sum_{f=1}^{N_f}\sum_{i=1}^\infty\left[h_{if}S_i^a[W]+
		\bar{h}_{if}S_i^{a,\dagger}[W]\right]\;,\nn
\eea
defining a 3d effective action by integration over the spatial link variables. 
The $S_i^{s,a}[W]$ depend on temporal Wilson 
lines $W(\bx)=\prod_{\tau=1}^{N_\tau}U_0(\bx,\tau)$, 
and the $S^s_i$ are $Z(N_c)$-symmetric while the $S^a_i$ are not.
The couplings of the effective theory are functions of the temporal extent $N_\tau$ of the 4d lattice, 
the fundamental representation character coefficient $u(\beta)=\beta/18+O(\beta^2)$ with lattice gauge coupling $\beta=2N_c/g^2$ and the hopping parameters $\kappa_f$, which for heavy quarks are 
related to the quark masses as $\kappa_f=\exp(-am_f)/2$.
Moreover, $\bar{h}_{if}(\mu_f)=h_{if}(-\mu_f)$.
The couplings are then ordered by increasing powers of their 
leading contributions. 
Up to several non-trivial orders, the gauge sector is dominated
by the nearest-neighbor interaction between Polyakov loops $L_i=\text{Tr}W(\bx_i)$, 
\bea
e^{-S_\text{eff}^s[W]}&=&\prod_{<ij>}[1+2\lambda \text{Re}L_iL^*_j]\;; \\
\lambda(u,N_{\tau}\geq5)&=&u^{N_\tau}\exp\bigg[N_{\tau}\bigg(4u^4+12u^5-14u^6-36u^7\nonumber\\
		&&\hspace*{-0.8cm}
		+\frac{295}{2}u^8+\frac{1851}{10}u^9+\frac{1055797}{5120}u^{10}+\ldots\bigg)\bigg]\;.
	\label{eq_lambda}
\eea
The convergence properties of the existing terms as well as explicit comparison with full 
4d thermal simulations demonstrate that for $\beta\lsim6.5$ the pure gauge sector is under 
control and the effect of higher couplings is negligible for $N_\tau\gsim 6$ \cite{efft_ym}.
When fermions are present, $\beta$ is shifted by $O(\kappa^4)$ corrections, which we neglect here. 

The $Z(N_c)$-breaking terms can be written as factors 
\beq
e^{-S^a_\text{eff}[W]}=\prod_n\Delta_n[W]\;.
\eeq
Summing all windings of the temporal loops this reads
\bea
\label{deltas}
\Delta_1&=&\prod_{f,i}\det[1+h_{1f}W_i]^{2}[1+\bar{h}_{1f}W_i^\dag]^{2}\;;\\
\Delta_2&=&\prod_{f,<ij>}\left[1-h_{2f}N_\tau\text{Tr}_c\frac{W_i}{1+C_fW_i}\text{Tr}_c\frac{W_j}{1+C_fW_j}\right]^2\;,\nn
\eea
with the couplings
\bea
h_{1f}&=&C_f\left[1+6\kappa_f^2N_\tau \frac{u-u^{N_\tau}}{1-u}+\ldots\right]\;;\nn\\
h_{2f}&=&C_f^2\frac{\kappa_f^2}{N_c}\left[1+2 \frac{u-u^{N_\tau}}{1-u}+\ldots\right]\;,
\eea
$C_f\equiv(2\kappa_f e^{a\mu_f})^{N_\tau}=e^{(\mu_f-m_f)/T}$, 
$\bar{C}_{f}(\mu_f)=C_{f}(-\mu_f)$.
From now on we consider $N_f=1$ and drop the index ``$f$''  (i.e.~$\mu=\mu_B/3$), 
which is sufficient to see the essential features.  
Finally we need meson and baryon masses, 
\bea
am_M&=&-2\ln(2\kappa)-6\kappa^2-24\kappa^2\frac{u}{1-u}+\ldots\;,\nn\\
am_B&=&-3\ln (2\kappa)-18\kappa^2\frac{u}{1-u}+\ldots\;.
\label{hadron}
\eea

Let us begin our analysis of the cold and dense regime with 
the combined static and strong coupling limit. 
In this case we have 
$\beta=\lambda=0$ and the partition function factorizes into exactly solvable single site
integrals:
\bea
\label{onesite}
Z(\beta=0) &=&
\Bigg[ \int \de W \left(1+CL+C^2L^\ast+C^3\right)^2\\
\hspace*{-25cm} & & \times \left(1+\bar{C}L^\ast+\bar{C}^2L+
		\bar{C}^3\right)^2\Bigg]^{N_s^3}=Z_1^{N_s^3}\;.\nn
\eea
The group integration only yields non-zero results if the trivial representation is contained in the 
products of loops. This results in the survival of hadronic degrees of freedom only,
\bea
Z_1&=&\left[1+4C^3+C^6\right]+2C\left[2+3C^3\right]\bar{C}\nn\\
&&\hspace*{-1.5cm}+\,2C^2\left[5+3C^3\right]
\bar{C}^2+2\left[2+10C^3+2C^6\right]\bar{C}^3\nn\\
&&\hspace*{-1.5cm}+\,2C\left[3+5C^3\right]\bar{C}^4+
2C^2\left[3+2C^3\right]\bar{C}^5\nn\\
&&\hspace*{-1.5cm}+
\left[1+4C^3+C^6\right]^3\bar{C}^6\;\stackrel{T\rightarrow 0}{\longrightarrow}\;\left[1+4C^{N_c}+C^{2N_c}\right]\;.
\eea
We recognize the partition function of an ideal gas of 
baryons ($\sim C^3$),
mesons ($\bar{C}C$) and composites of those, as already discussed in \cite{lp}. 
For finite chemical potential and zero temperature, $\bar{C}\rightarrow 0$, we are left with 
baryons and have reinstated $N_c$ to illustrate 
the meaning of the exponents. 
Prefactors are identified as spin degeneracy,
i.e.~we have a spin-3/2 quadruplet for the
three-quark baryon and a spin zero baryon made of six quarks.

The quark density is now easily calculated,
\beq
n=
\frac{T}{V}\frac{\partial}{\partial \mu}\ln Z=\frac{1}{a^3}\frac{4N_cC^{N_c}+2N_cC^{2N_c}}{1+4C^{N_c}+C^{2N_c}}\;.
\eeq
In the high density limit the expression reduces to
\beq
\lim_{\mu\rightarrow\infty}(a^3n)=2N_c \equiv N_c(a^3n_{B,\mathrm{sat}})\;.
\eeq 
As required for fermions obeying the Pauli principle, the quark density in lattice units saturates
once all available states per lattice site labeled by spin, color (and flavor for $N_f>1$) are occupied.
Note that summation over all
windings of the Wilson lines is necessary in order to obtain a determinant in the form \eq(\ref{deltas}),
while a truncation to finite order would not show saturation. 
Next, consider finite chemical potential and the zero temperature limit,
\bea
\lim_{T\rightarrow 0} a^4f&=&\left\{\begin{array}{cc} 0, & \mu<m\\
	2N_c (am-a\mu), & \mu>m\end{array}\right.\;;\nn\\
\nn\\
\lim_{T\rightarrow 0} a^3n&=&\left\{\begin{array}{cc} 0, & \mu<m\\
	2N_c, & \mu>m\end{array}\right.\;.
\eea                                      
Thus the static strong coupling  
limit shows the silver blaze property, with zero quark density for $\mu<m$ and 
a jump to saturation density for $\mu>m$, corresponding to a first order phase transition at
quark chemical potential $\mu_c=m$. In the static strong coupling limit the baryon mass
is $am_B=-3\ln(2\kappa)=3am$, i.e.~the onset happens at $\mu_B=m_B$ and
satisfies the bounds in \cite{cohen2}. (For static quarks, $m_B/3= m_\pi/2$).
In the dense phase the free energy scales with $N_c$, 
consistent with the conjecture
in \cite{quarky}. For $T\neq 0$ the step function is 
smeared out and the transition becomes smooth, 
as expected for a non-interacting system.
 
Next we consider the interacting theory to $O(\kappa^2)$
at finite gauge coupling and quark mass. 
In this case the partition function has to
be computed numerically and for finite values of $N_\tau$, i.e.~the 
zero-temperature limit has to be approached numerically. 
The effective theory features a sign problem, which however is mild compared
to that of the full 4d theory and can be overcome by reweighting 
methods using a standard Metropolis
algorithm. For $h_{2f}=0$ the effective theory can  
be cast into a flux representation and simulated
with the worm algorithm, without any sign problem. The two approaches gave consistent results for
the deconfinement transition at finite temperature and density \cite{efft_kappa}.
Here we include $h_2\neq 0$, as it is the first coupling of quark-quark terms $\sim L(\bx)L(\by)$. 
In this case we could not find a flux representation free of the sign problem and instead employ complex Langevin simulations as an independent check. Indeed, that algorithm
has been shown to work for a simple $SU(3)$ one-site model as well as QCD in the heavy dense
limit \cite{aarts2}, which have structures very similar to our effective theory. All our simulations
satisfy the convergence criterion in terms of the Langevin operator specified in \cite{etiology}.

In order to reach continuum QCD, we work at small hopping
parameters $\kappa\lsim 10^{-3}$, close to but not in the static limit. 
In this case we can use the non-perturbative beta-function of pure gauge theory 
for the lattice spacing in units of the Sommer parameter, 
$a(\beta)/r_0$ with $r_0=0.5~{\rm fm}$ \cite{sommer}.
Moreover, near the static limit 
\eq(\ref{hadron}) gives a good approximation to the hadron masses. 
Finally, temperature is tuned
via $T=(aN_\tau)^{-1}$. 
To begin, let us consider $T=10$ MeV and
$m_\pi=20$ GeV. Because of the short Compton wave length, 
small lattices are sufficient for baryonic quantities, 
with negligible differences between $N_s=3,6$.
Once the lattice spacing is chosen, $\beta$ is fixed and 
\eq(\ref{hadron}) determines the corresponding $\kappa(\beta)$.

\begin{figure}[t]
\vspace*{-0.2cm}
\includegraphics[width=0.32\textwidth]{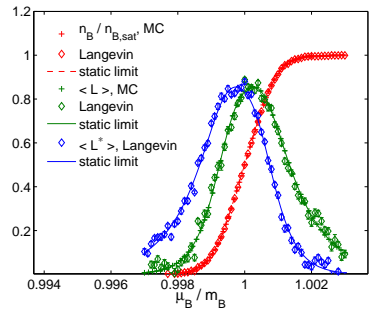}
\vspace*{-0.3cm}
\caption[]{
Baryon density $n_B / n_{B,\mathrm{sat}}$, Polyakov loop $\langle L \rangle$ and conjugate Polyakov loop 
$\langle L^\ast \rangle$ as a function of $\mu_B/m_B$ obtained from Monte Carlo calculations ($N_s = 3$), complex Langevin ($N_s=6$) and the static strong coupling limit, respectively. 
Lattice parameters $\beta=5.7, \kappa = 0.0000887, N_\tau=116$ correspond to $m_M=20$ GeV, 
$T=10$ MeV, $a=0.17$ fm.}
\label{raw}
\end{figure}

\fig\ref{raw} shows the baryon density in lattice units as 
a function of chemical potential in units of the baryon mass for $\beta=5.7$. 
It is consistent with zero until the chemical
potential approaches $m_B/3$, where a transition or crossover is clearly visible 
which quickly reaches saturation level. The rise in the baryon density is accompanied by
a rise in the Polyakov loop. 
This feature was also seen in 4d Langevin simulations \cite{aarts2} 
and in the chiral strong coupling limit in the staggered discretization \cite{chiral}.
By contrast, in two-color QCD with lighter masses Bose condensation and the rise of the Polyakov loop appear to reflect two distinct transitions \cite{hands}. 
It is not clear to us whether the rise of the Polyakov loop signals deconfinement in the presence of
matter. Evaluating $\langle L^\ast\rangle,\langle L\rangle$ using \eq(\ref{onesite}) with $\bar{C}=0$, 
the Polyakov/conjugate loop gets screened by the  third/second terms without changing the nature of the medium, which is hadronic. This also explains why $L^\ast$ is screened before $L$ when $\mu>0$,
while the opposite happens for $\mu<0$. 
The ensuing decrease is a consequence 
of saturation: all color orientations get populated once the lattice approaches filling. 
All quantities in \fig\ref{raw}  
agree between the Metropolis and Langevin algorithms, 
the latter is vastly superior on larger volumes. 

It is very striking that the numerical results are reproduced excellently by the analytic solution to the 
free, static hadron gas discussed earlier. That the static limit works well is easy to understand,
since our quarks are exceedingly heavy and $O(\kappa^2)$ corrections 
are tiny. What is less obvious is that a simulation at $\beta=5.7$
is well approximated by the strong coupling limit, $\beta=0$. The reason is that the effective
coupling of the gauge sector $\lambda(\beta=5.7,N_\tau=115)\sim10^{-27}$. This is an important 
observation. The convergence of the strong coupling expansion 
is sufficiently fast to allow for an accurate estimate of the convergence radius
$\beta_c<6$ for $N_\tau\leq16$ in \cite{efft_ym}. For $\beta$ in the same range, 
lowering temperature increases $N_\tau$ and thus improves convergence in two ways: 
we move away from the limiting convergence radius and
$u(\beta)<1$ gets suppressed by ever higher powers. In other words, cold QCD is more amenable to 
the strong coupling expansion than hot QCD, and the pure gauge sector plays a negligible role
for the dynamics.

\begin{figure}[t]
\vspace*{-0.3cm}
\includegraphics[width=0.32\textwidth]{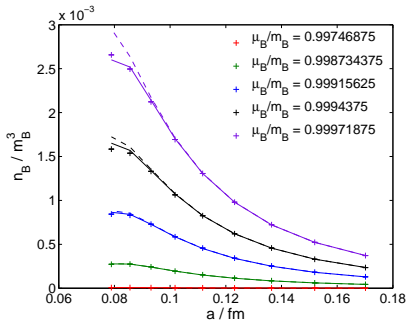}
\vspace*{-0.3cm}
\caption[]{Baryon density $n_B/m_B^3$ as a function of the lattice spacing $a$ at $T = 10$ MeV and several values of the chemical potential. Crosses correspond to MC data $(N_s=3)$, lines to the static strong coupling limit, Eq.(\ref{deltas}), (dashed) and including $\mathcal{O}(\kappa^2)$-corrections (solid).}
\label{scaling}
\end{figure}
Simulations of the effective theory being cheap, we have computed the baryon density for 
nine gauge couplings $5.7<\beta<6.1$, corresponding to lattice spacings
$0.17\text{ fm} > a > 0.07\text{ fm}$. The scaling of the result in physical units is shown in 
\fig\ref{scaling}. Since the quark density is a derivative of the physical pressure with respect to an external parameter, it is a finite quantity that does not renormalize in a non-perturbative
calculation. (The pressure requires subtraction of divergent vacuum energies, 
$p_\text{phys}(T)=p(T)-p(0)$; however, these are $\mu$-independent 
and $\partial_\mu p_\text{phys}$ is finite.) Massive Wilson fermions have $O(a)$ lattice corrections,
hence the continuum approach is
\beq
\frac{n_{B,\text{lat}}(\mu)}{m_B^3}=\frac{n_{B,\text{cont}}(\mu)}{m_B^3}+A(\mu) a +B(\mu) a^2+\ldots
\label{extra}
\eeq  
This behaviour is borne out by the data for $a>0.09$ fm in \fig\ref{scaling}. On the other hand, for 
$a\rightarrow 0$ we see a downward bend that violates scaling and signals that our truncated
series in $\beta,\kappa$ are no longer valid:
as the lattice gets finer, $\beta$ and $\kappa(\beta)$ grow and our effective action
eventually must fail. Adding or removing $O(\kappa^2)$ corrections indeed affects this 
downward bend, but not
the rest of the curve. Note that there is a trade-off between $\kappa$ and $\beta$. The lighter we make 
the quark mass, the larger $\kappa$ for a given lattice spacing and the earlier the breakdown of the
hopping expansion. 
Thus the scaling behavior of the baryon density tells us when
our effective theory breaks down. 

\begin{figure}[t!!!]
\vspace*{-0.2cm}
\includegraphics[width=0.35\textwidth]{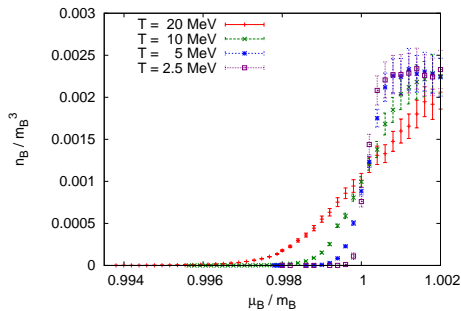}
\vspace*{-0.4cm}
\caption[]{Baryon density in the continuum. 
In the zero temperature limit a jump to nuclear matter builds up.}
\label{blaze}
\end{figure}
For the very heavy quarks studied here, our series are controlled for 
lattice spacings down to $a\gsim 0.09$ fm, which is just entering the regime with leading order lattice 
corrections. Cutting our data for $a<0.09$-$0.11$ fm, we perform continuum extrapolations based
on five to seven lattice spacings by fitting to \eq(\ref{extra}).
We have followed this procedure for four different temperatures, 
resulting in the baryon densities
in \fig\ref{blaze}. Clearly, the silver blaze property and a jump in
baryon density get realized also in the interacting, dynamical theory
as temperature approaches zero. 
Interestingly, the saturation density beyond onset, 
when expressed in units of $m_B$, is of the same order of magnitude
as the physical nuclear density 
$\sim 0.16$ fm$^{-3}\approx 0.15 \cdot 10^{-2} \;m_\text{proton}^3$.
Finite size analyses using $N_s=3,4,6,8$ show that the onset at $T=2.5$ MeV
is still a smooth crossover, i.e.~$T$
is too high for a first order transition. 
Presently $T$ cannot be drastically reduced because 
$\kappa^2$-corrections to the determinant get enhanced $\sim N_\tau$,
\eq(\ref{deltas}). 
For physical quark masses the onset transition persists up to $T\sim 10$ MeV.
More work is needed to study whether this difference is due to the larger
quark mass or to the truncation of the hopping series.

For light quarks our truncated hopping series is not 
reliable. 
How far the series can be extended and 
whether the pure gauge sector is similarly suppressed
remains to be seen. 
We plan to address $\kappa^4$-corrections and 
details of 
the Langevin simulations in a future publication.

\noindent
{\bf Acknowledgements:}
This project is supported by the German BMBF, contract number 06MS9150 and by the Helmholtz International Center for FAIR within the LOEWE program launched by the State of Hesse.


%
\end{document}